\documentclass[referee]{raa}            
\usepackage{graphicx,times}             

\begin{document}

 \title{Mass flow in the circumbinary disk with gap around supermassive binary black holes
 }


   \author{Ning-Yu Tang
          \inst{1}
   \and
          Ye-Fei Yuan
          \inst{1}
          }

   \institute{Key Laboratory for Research in Galaxies and Cosmology CAS,Department of Astronomy, University of Science
and Technology of China, Hefei, Anhui, 230026, China; {\it yfyuan@ustc.edu.cn}\\
      }

   \abstract{In this paper, we study the interaction between the supermassive binary black holes in elliptical orbit and their surrounding disk with a gap.
 The gap in the disk is a low density region formed due to the tidal effects
of the less massive black hole. The binary we have investigated has a sub-parsec separation and is coplanar with
the disk. We find that the maximum variation of the surface density in the gap reaches $50\%$ during an orbital period. However, in other regions of the disk, the density variation is much less than $1\%$.
Furthermore, we calculate the corresponding variation of spectral energy distribution within a period, but little variation is found.  The reason for these results is that the viscosity timescale of the disk at the binary radius is much longer than the orbital period of the binary.
        }

   \maketitle

\section{Introduction}
\label{intro}

 Supermassive binary black holes (SMBBHs) is believed to form during galaxy mergers which happens frequently in the universe according to the hierarchical models of galaxy formation.  It is widely accepted that the merging of the two SMBHs would go through three stages (Begelman et al.~\cite{Bege80}; Colpi et al.~\cite{Colp11}). The standard picture is as follows. After the initial phase of merging of the galactic cores due to the dynamical friction, the two SMBHs would form a bound pair. The second phase in which the SMBBHs hardens via 3-body scattering off single star is still uncertain. It is of theoretical difficulty to shrink the separation of a SMBBHs by a factor of $\sim 100$ after its formation at a separation of $\sim 1$pc ( Merritt et al.~\cite{Merr02}). This is known as the ``final parsec problem".  At the final phase, when SMBBHs reaches a separation of  $\sim 0.01$pc, the two black holes will coalesce within a relatively short time interval through emission of gravitational waves (GWs) (Sathyaprakash \& Schutz et al.~\cite{Sath09}).
    Measuring the separation distribution between pairs of SMBHs helps to constrain the timescales of the different stages of SMBHs merging, and to estimate the number of gravitational wave sources that can be detected with the Laser Interferometer Space Antenna (LISA) in one year (Dotti et al.~\cite{Dott06}).

 Currently, there is no direct detection of GWs, the search for SMBBHs  mainly focus on the first and second phase of the merging SMBHs (Komossa et al.~\cite{Komo06}; Komossa et al.~\cite{Komo08}; Colpi et al.~\cite{Colp11}). Some SMBBHs, for instance, NGC6240 (Komossa et al.~\cite{Komo03}), COSMOS J100043.15+020637.2 (Comerford et al.~\cite{Come09}) and J0402+379 (Rodriguez et al.~\cite{Rodr06}), have been identified spatially in a single galaxy via X-ray, optical and radio imaging spectroscopy.  A few galaxies which contain double-peaked [O III] or H$\beta$ lines with velocity offsets are suggested as pairs of AGNs ( Zhou et al.~\cite{Zhou04}; Dotti et al.~\cite{Dott09}; Wang et al.~\cite{Wang09}; Shen et al.~\cite{Shen10}; Liu  et al.~\cite{Liux10}). In addition, several other ways have been proposed to identify SMBBHs: X-shaped morphology of radio lobes (Merritt et al.~\cite{Merr02}),  double-double radio galaxies (Liu  et al.~\cite{Liuf03}), orbital motion of the compact core with a periodic flux variation (Sudou et al.~\cite{Sudo03}), periodic optical and radio outbursts (e.g., OJ 287) (Valtonen et al.~\cite{Valt08}), and so on.

 Eccentricity of SMBBHs in merging galaxies with large amount of gas is proved to be high (Roedig et al.~\cite{Roed11}), which leads to periodic accretion flows onto the SMBH and then periodic occurrence of peaked flares in light curves (Hayasaki et al.~\cite{Haya07}).  Hayasaki~(\cite{Haya12}) carried out the smooth particle hydro-dynamics simulation on circumbinary disk with a central cavity and calculated the periodic mass accretion rate $\dot{M}$ onto the SMBHs. The corresponding periodic  luminosity (luminosity $L= \eta \dot{M}c^2$) is believed as the indicator of SMBBHs at sub-parsec separation. But the mass accretion rate in their work is defined as the mass rate falling into an fixed ``accretion" sphere without considering the detailed accretion physics. The radius of the fixed ``accretion" sphere is chosen to be $0.1a$, where $a$ is the semi-major axis of the binary orbit.

  In this paper, the matter is accreted  through a disk existing around the primary SMBH, rather than a defined "accretion" sphere. The matter across the gap flows onto the inner disk and is accreted into the primary SMBH by viscosity effect. This is a more realistic and physical consideration.
 
The outline of this paper is as follows.
 In section 2, we  introduce the basic physics of
binary system, followed by the description of the simulation in section 3.
 Finally, results and discussions are presented in section 4.

\section{Physical model of the accretion in SMBBHs}
\label{setup}
\subsection{Basic equations governing circumbinary disk}
 It is assumed that the secondary SMBH  exists in the accretion disk of the primary SMBH and its orbital plane is coplanar with the  disk plane (Bogdanovic et al.~\cite{Bogd07}; Dotti et al.~\cite{Dott10}).
The dynamics of this system is similar to that of the protoplanetary disk
in the preliminary formation of a star. The latter has been widely discussed
in the past decades (Goldreich et al.~\cite{Gold80}; Takeuchi et al.~\cite{Take96}; Espaillat et al.~\cite{Espa08}). The  hydrodynamic equations of the evolution of
the proto-planetary disk are described as follows (Kley et al.~\cite{Kley99}; Crida et al.~\cite{Crid09}):
\begin {equation}
\frac{\partial \Sigma}{\partial t} + \nabla\cdot(\Sigma {\bf u} )
                =  0,  \label{Sigma}
\end{equation}
\begin {equation}
 \frac{\partial (\Sigma \upsilon)}{\partial t} + \nabla\cdot(\Sigma \upsilon {\bf u} )
  = \Sigma \, r ( \omega + \Omega)^2
        - \frac{\partial p}{\partial r} - \Sigma\frac{\partial \Phi}{\partial r} + f_{r},
      \label{u_r}
\end{equation}
\begin {equation}
 \frac{\partial [\Sigma r^2 (\omega + \Omega)]}{\partial t}
   + \nabla \cdot[\Sigma r^2 (\omega + \Omega) {\bf u} ]
         =
        - \frac{\partial p}{\partial \varphi} - \Sigma \frac{\partial \Phi}{\partial \varphi}
   + f_{\varphi},
      \label{u_phi}
\end{equation}

 These equations are shown in cylindrical coordinates ($r,\varphi,z$), where $r$ is the radial coordinate, $\varphi$ is the azimuthal angle, $z$ is the vertical axis.  The reference frame rotates with the angular velocity $\Omega$. The rotation is around the $z$-zxis, i.e. $\vec{\Omega}=(0,0,\Omega)$. The disk locates in the $z=0$ plane.

 Equation (1) represents the conservation of mass. $\Sigma$ denotes the surface density and $\bf u$ is the velocity vector of the fluid.
Equation (2), (3) represent the momentum equation in $r$ and $\varphi$ direction.  The symbol $\upsilon = u_r$ means radial velocity and $\omega = u_\varphi/r$ means the angular velocity of the flow, both of which are measured in the corotating frame. $p$ is the vertically integrated (two-dimensional) pressure.
The gravitational potential $\Phi$ is contributed by both the primary SMBH
with mass $M_p$ and the secondary BH with mass $M_s$, which is given by
\[
    \Phi = - \frac{G M_p}{| {\bf r} - {\bf r}_p |}
       \left( 1 + \frac{M_s | {\bf r} - {\bf r}_p |}{M_p | {\bf r} - {\bf r}_s |}\right)
       = \Phi_0 (1+\delta),
\]
where $G$ is the gravitational constant and ${\bf r}_p,
{\bf r}_s$ are the position vectors of the primary and secondary
BH, respectively. $\Phi_0$ is the gravitational potential induced by the primary SMBH and $\delta$ represents the potential perturbations induced by the secondary BH. If $\delta$ is small, Fourier decomposition of
the time-periodic gravitational potential of the binary is applicable. The potential is given by a double series
\[   \Phi(r,\theta,t) = \sum_{ml} \phi_{ml}(r) e^{[i(m\theta - l\Omega_B t)]},
\]
where $l$ is the time-harmonic number, $m \geq 0$ is the azimuthal number, $\phi_{ml}(r)$ is the radially variable potential component,
  \[ \phi_{ml}= \frac{1}{2\pi^2}\int_{0}^{2\pi}\int_{0}^{2\pi}\Phi \cos(m\theta-l\Omega_Bt)d\theta d(\Omega_B t),
  \]
  and $\Omega_B = (GM/a^3)^{1/2}$ is the mean motion of the binary, with $M$ and $a$ denoting the total mass and semimajor axis of the binary, respectively (Artymowicz et al.~\cite{Arty94}). Two terms in equation (2)-(3), $\frac{\partial \Phi}{\partial r}$ and $\frac{\partial \Phi}{\partial \varphi}$, act as tidal effects on the disk. $f_{r}$ and $f_{\varphi}$ represent the viscous force per unit area acting in the radial and azimuthal direction. The explicit forms of $f_{r}$ and $f_{\varphi}$ are given by Kley~(\cite{Kley99}).

\subsection{Resonance}

  Artymowicz~(\cite{Arty94}) had analytically discussed the gravitational interaction of a generally eccentric binary star system with the circumbinary gaseous disks. The interaction between the circular binary and the circumbinary disk can occur via $(m,l)=(1,1)$ resonance (see section 2.1 for the meaning of $m, l$) and the higher
noneccentric resonances like $(m,l)=(2,2)$, while the eccentric binary
will induce resonances like $(m,l)=(2,1),(3,1),(4,1)$ or higher harmonics. The resonance would lead to the tidal torque.
When the total tidal torque  at the edge of the disk is lager than the viscous torque,
 a gap is opened in the disk (Lin \& Papaloizou et al.~\cite{Lin86b}).
The gap-opening condition for $(1,1)$ circular resonance is given by,
\begin{equation}
\alpha^{1/2}\left(\frac{H}{r}\right)\leq0.24q(1-q)(1-2q)~~~.
\label{hilldef}
\end{equation}
\noindent

Here, $q$ is the mass ratio between the SMBBHs. The circumbinary disk around the SMBBHs can be described  by $\alpha$ prescription ( Shakura \& Sunyaev et al.~\cite{Shak73}). $r$  is the radius to the central mass point and $H$ is the corresponding scale height of the disk at radius $r$.

For the eccentric binary, the $(m,l)=(2,1),(3,1)$ resonance become more important.
The gap-opening condition for $(2,1), (3,1)$ resonance is given by
\begin{equation}
\alpha^{1/2}\left(\frac{H}{r}\right)\leq1.76eq(1-q)~~~
\label{hilldef}
\end{equation}
and
\begin{equation}
\alpha^{1/2}\left(\frac{H}{r}\right)\leq1.42e^2q(1-q)(1-2q)~~~
\label{hilldef}
\end{equation}
respectively. Here
$e$ is the eccentricity of the binary.
 If the values of $\alpha$, $\frac{H}{r}$ and $q$ are taken to be 0.05, 0.01 and 0.01, the minimum eccentricity to open the $(2,1)$ and $(3,1)$ resonance
gap is about $0.127$ and $0.397$, respectively.

\subsection{Spectral energy distribution of the disk}

The disk is assumed to be in the local thermal equilibrium, that is,
the heating and cooling are locally balanced at each point of the disk, $Q_{+} = Q_{-}$.
The viscous heating term reads $Q_+=9\Sigma\nu\Omega^2/4$, where $\nu$ is the kinetic viscosity. The cooling term is given by the thermal emission:  $Q_- =2\sigma_R T_{eff}^4$, where $\sigma_R$ is the Stefan-Boltzmann constant,
$T_{eff}$ is the effective temperature of the disk. So the effective temperature of each element of the disk is given by the equation:
\begin{equation}
\ T_{eff}(r,\varphi)^4=\frac{9}{8}\frac{\nu}{\sigma}\Sigma(r,\varphi)\frac{GM}{r^3}
\end{equation}
\noindent

The SED from an thermally emitting disk with the
temperature as a function of the radius can be written as:
\begin{equation}
F_\lambda=\int_{0}^{2\pi}\int_{R_\mathrm{in}}^{R_\mathrm{out}}B_\lambda[T_{eff}(r^\prime,\varphi^\prime)]g(r^\prime){r^\prime}dr^\prime{d\varphi^\prime},
\label{flambdased}
\end{equation}
where $B_\lambda(T_{eff})$ is the Planck function and $g(r^\prime)$ is a function
that describes the emissivity as a function of radius.  We assume that
$g(r^\prime)=1$  where the disk material is present and
$g(r^\prime)=0$ elsewhere.

\section{Physical model and parameters}
\label{sec:setup}
We carry out simulation using the  FARGO-ADSG code (Baruteau et al.~\cite{Baru08}) in a two dimensional (R,$\phi$) fixed polar grid. FARGO-ADSG is an extended version of FARGO (Masset et al.~\cite{Mass00}). In the FARGO-ADSG code, additional orbital advection which removes the average
azimuthal velocity has been applied so that the truncation error is reduced and the time-step allowed by the Courant-Friedrichs-Lewy condition is significantly increased. The code was initially used to simulate the tidal interactions of planet-disk. We adopt it because of the similarity between the planet-disk and the SMBBHs-disk.
The key model parameters are listed as follows:

\par\noindent\emph{Binary parameters---}
 The mass of the primary and secondary black hole is taken to be $M_1 =10^8 M_{\odot}$ and $M_2 = qM_1 = 4\times10^6 M_{\odot}$, respectively. This is common in galaxy-dwarf galaxy minor mergers whose mass ratio $q$ ranges from 0.001 to 0.1. In despite of the small  mass ratio, recent numerical simulations have shown that the local processes can lead these binaries to be close enough (Bellovary et al.~\cite{Bell10}). The initial semi-major axis $a_0$ and eccentricity $e_0$ of the binary is taken to be $10^3 r_{\rm g}$($\approx 10^{-2} pc)$ and $0.5$, respectively.
\par\noindent\emph{Disk parameters---}
Before its evolution, the disk is axisymmetric and rotates at Keplerian angular
frequency $\Omega(r)$. The disk extends from $r=10 r_{\rm g}$ to $r=4\times10^3 r_{\rm g}$
in our simulation. The mesh is $256\times156$ in the polar coordinates ($r,\theta$).
 The initial gas density of the disk is about $ 6.6\times10^5(r /a_0)^{-1/2} g\,cm^{-2} $, which means that the initial disk mass is about $ 0.1M_1$. The aspect ratio  $h = H/r$ is assumed as $0.02$ and remains constant throughout the simulation. In addition, the disk is modeled with a constant
kinematic viscosity $\nu=2\times10^{15} m^{2}/s$, which corresponds to an alpha viscosity $\alpha \approx 0.05$ at $0.01pc$. The criteria for  gap opening  is satisfied in our model.

The inner boundary condition in the code is chosen to be open such that material can flow
outside of the mesh, on its way to the central object. This results in the decrease of  disk mass with time. The outer boundary condition which is not independent of the inner boundary condition can not be selected in the code.

\begin{figure}
\begin{center}

  \includegraphics[angle=-0,width=5.0in]{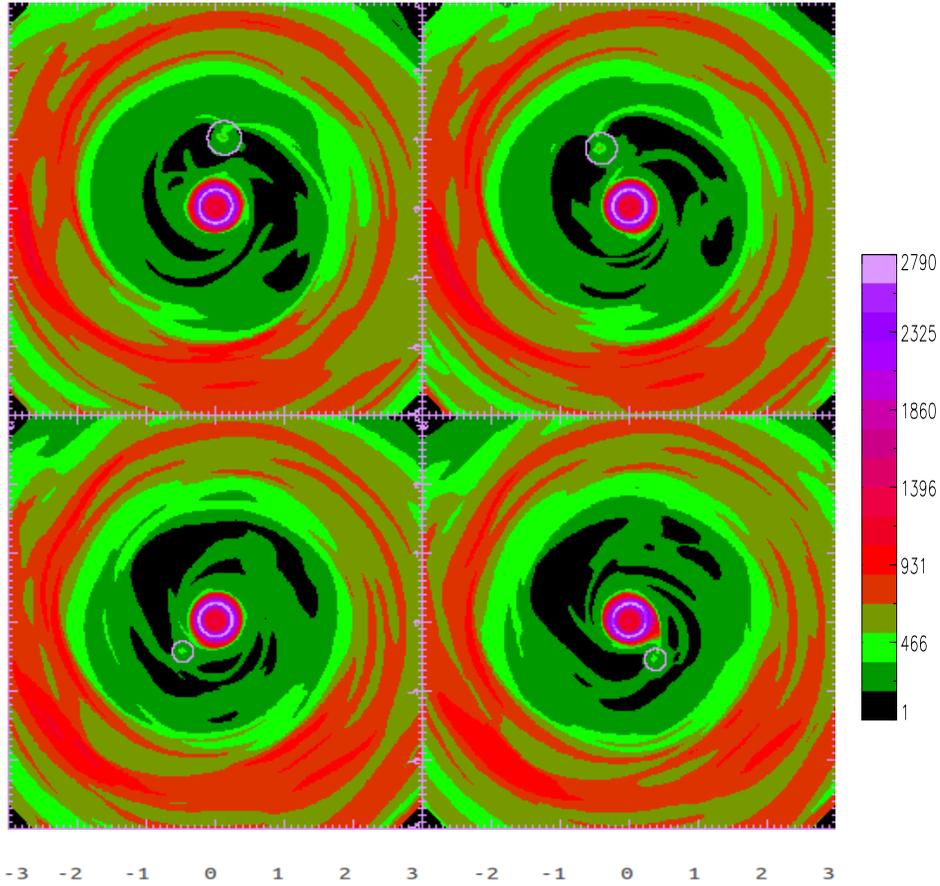}
 \end{center}
\caption{\label{fig:fig1} Two-dimensional distributions of mass density
at different orbital phases of the SMBBHs.
{\em Top left} panel shows the result at apocenter phase.
{\em Bottom right} panel shows the result at pericenter phase.
{\em Top right} and {\em Bottem left} panels show
two middle phases between the apocenter and pericenter phase, respectively.
The plum circles represent the position of secondary BH.
The primary black hole locates at the center, the coordinates are in units of
$1000R_{g}$ of the primary black hole.
The density levels of each panel can be seen in the color bar on the right side.
The black colour represents the minimum value of the surface density which is about $9.46\times10^3 g\cdot cm^{-2}$.
The plum represents the maximum value which is 2790 times of the
value of the black. The number of other colours shows the relative value of the surface density.
 }

\end{figure}


\begin{figure}
  \begin{center}
  \includegraphics[angle=-90,width=3.4in,totalheight=3.0in]{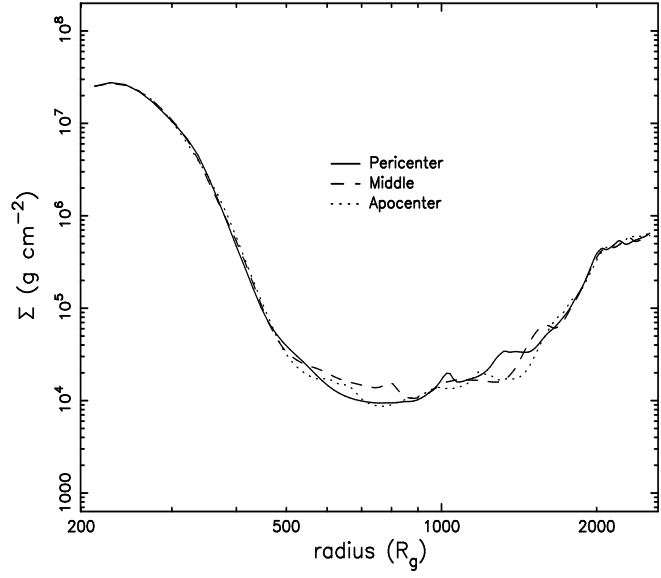}
  \end{center}
  \caption{\label{fig:fig2} One-dimensional distribution of the surface density
 within the radii between  $220R_g$ and $2600R_g$.}
\end{figure}


\begin{figure}
  \begin{center}
  \includegraphics[angle=-90,width=3.4in,totalheight=3.0in]{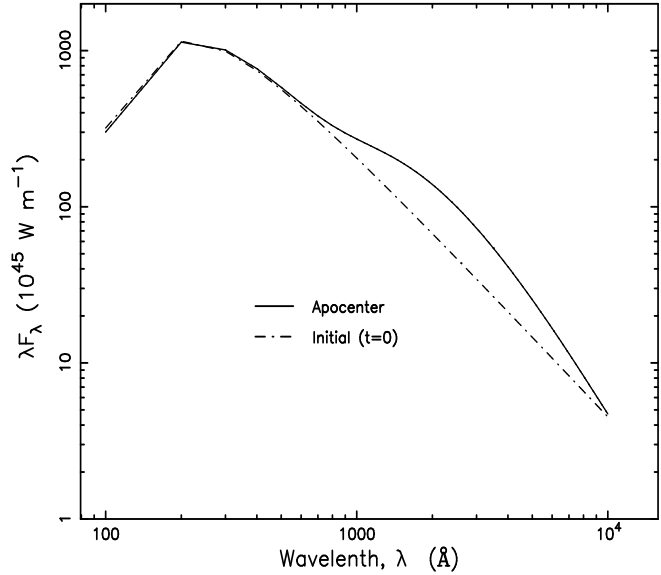}
  \end{center}
  \caption{\label{fig:fig3} Spectral energy distribution (SED) of the radiation
from the accreting flow in the system of SMBBHs. After a stable gap is opened, solid line represents the SED at apocenter. The curves of SEDs at other orbital phases (middle and pericenter) vary  little compared with that at apocenter and it is difficult to distinguish the curves at different orbital phases.  The curves of SEDs at other orbital phases (middle and pericenter) are not plotted in this figure.  For comparison, the initial SED is shown in the dot-dash line. At the initial moment, there is no gap
in the disk.}
\end{figure}

\section{Results and Discussions}
\label{result}
In this work, the simulation runs until a stable gap is opened in the circumbinary disk of SMBBHs.
 Our results are shown in Fig. 1-3.

We illustrate in Fig.1 the two-dimensional surface density distribution of the disk when the binary has a stable eccentricity ($e=0.3$). A gap is opened in the disk. The secondary SMBH  causes the gas flow across the gap  when it moves from the apocenter to the pericenter. Furthermore, a ring of high density exists in the inner disk, that can be attributed to the matter accumulation.

In order to clearly show the variation of the surface density, one-dimensional distribution of the surface density
in one orbit is shown in Figure 2. The surface density  varies significantly
in the region of the gap (about $400 R_g \sim 1600 R_g$). The variation reaches $50 \%$ at $800 R_g$. However, the total amount of the gas crossing the gap is still too low to
 change the surface density of the inner disk (less than $1 \%$ below $400 R_g$).

It is meaningful to discuss the potential observational phenomena.
Based on Equation (7) and (8), the emission fluxes at the various wavelengths are shown in Fig.3.  There is little variation of the SED
among the different orbital phases of the binary.

  In this work we simulate the interaction between the eccentric SMBBHs and their circumbinary disk. Surface density of the disk evolves under the action of binary torques in one orbital period. The orbital period is $\sim 10 yr$ in our model. For comparison, we estimate the timescale of viscosity (therefore the accretion timescale) near the outer edge of the inner disk (about $400R_g$). The timescale of viscosity is
$t_{visc} \sim R^2/\nu \sim 2.28\times10^5 yr$ with our model parameters. It is obvious that the accretion timescale is much longer than
the orbital period. Therefore, the gas across the gap only accumulates at the outer boundary of the inner disk. It is impossible for the accumulated gas to be accreted into the central SMBH in one orbital period. Consequently, the surface density of inner disk does not vary  significantly, as shown in Fig.2. The effective temperature in the disk is $T_{eff} \propto \Sigma^{1/4}/r^{3/4}$, so
the effective temperature of the inner disk with larger surface density and smaller radius is higher than that of the other region. The emission from the inner disk with little variation dominates the thermal radiation of the disk. Therefore, the SED of the disk changes little as shown in Fig.3.

It is a natural idea to reduce the timescale of viscosity
by increasing the value of the kinetic viscosity $\nu$.
However, as we argue below, this will be contradictory to the
criteria of the opening of a gap.
Equation (5) must be satisfied to open a gap in the disk if we take (2:1) resonance gap as an example,
which consequently results in a constraint on  $\alpha$.
 The kinetic viscosity is $\nu=\alpha h^2 a^2 \Omega_{k}$, with the help of equation (5), we get $\nu\leq 3 e^2 q^2 (1-q)^2 a^2 \Omega_{k}$.
Applying the typical parameters in our work ($e=0.3,q=0.04,a=0.01 pc,\Omega_{k}= \sqrt{GM_p/a^3}=2.21\times10^{-8} rad/s $ ),
the kinetic viscosity at the binary radius satisfies $\nu\leq 8.55\times 10^{17} m^2/s$.
Therefore, the typical viscosity timescale at the orbital radius ($t_{visc}\sim R^2 / \nu \sim 3000 yr$) is
at least $300$ times longer than that of the binary period ($\sim 10 yr$).

Recently, Bon~(\cite{Bon12}) has shown that a SMBBHs exists in the core of Seyfert galaxy NGC 4151.  The periodic variations in the light curves and the radial velocity curves of NGC 4151 can be accounted for by an eccentric ($e=0.42$), sub-parsec Keplerian orbit with a 15.9 year period. The periodic variations in the observed $H_\alpha$ line shape and flux are well explained in the model of the shock which is generated by  the supersonic motion of the SMBBHs through the surrounding medium, rather than the periodic disk accretion.
As we have argued that the periodic accretion of eccentric binary with sub-parsec separation does not cause obvious SED variation, this is consistent
with  the result of Bon~(\cite{Bon12}).
As we adopt the $\alpha$ prescription of the disk in this work, it needs to be checked whether this prescription is still valid for the closed SMBBHs system. We hope that the future observations of SMBBHs candidates at sub-parsec separation will provide more information to constrain the properties of their accretion disks.

In this paper, the effect of GWs is neglected. We discuss the reason below. The timescale of orbital decay of the SMBBHs through GWs emission (Peters et al.~\cite{Pete64}) is
\[t_{GW} \simeq 10^4\frac{(1+q)^2}{q}\left(\frac{10^8M_\odot}{M_{\rm total}}\right)^3(\frac{a}{2\times10^{-3}pc})^4\frac{1}{f(e)} yr.
\]
 Here, $f(e)=(1+\frac{73e^2}{24}+\frac{37e^4}{96})(1-e^2)^{-7/2}$, $e$ is the eccentricity of the binary system. $t_{GW}$ is about $8.8\times10^7 yr$ in our model and is much longer than the viscous timescale ($t_{vis}\simeq 10^6 yr$) at the orbital radius. Therefore it is reasonable to neglect the effect of GWs  because of the domination of viscous effect.

 The disk is modeled as a two-dimensional ($r,\varphi$) system, by using the vertically averaged quantities. Two main arguments lie behind this choice. First, on a physical basis, the Hill radius of a massive object is large or comparable to the disk semi-thickness. Second, a less massive secondary SMBH has a weak impact on the disk, requiring a higher resolution to compute properly and highlight its effects. In the 2D simulation, it is more easy to increase the resolution of numerical simulation than in the 3D one.  D'Angelo~(\cite{D'Ang03}) showed that two-dimensional simulations still yield reliable results when the mass ratio $q\geq 10^{-4}$. When the mass ratio $q < 10^{-4}$, three-dimensional simulations are needed. The mass ratio is about 0.04 ($q=0.04$) in our work, so the result might be still reliable.

\begin{acknowledgements}
We would like to thank Prof. J.X. Wang for helpful discussions.
This work is partially supported by
National Basic Research Program of China (2009CB824800, 2012CB821800),
the National Natural Science Foundation (11073020, 11133005, 11233003),
and the Fundamental Research Funds for the Central Universities (WK2030220004).
\end{acknowledgements}

\end{document}